\documentclass{article}

\usepackage{aimc2025}
\usepackage{natbib}
\usepackage[utf8]{inputenc} %
\usepackage[T1]{fontenc}    %
\usepackage{hyperref}       %
\hypersetup{
  colorlinks   = true, %
  urlcolor     = blue, %
  linkcolor    = black, %
  citecolor   = black %
}
\usepackage{url}            %
\usepackage{booktabs}       %
\usepackage{amsfonts}       %
\usepackage{nicefrac}       %
\usepackage{microtype}      %
\usepackage{graphicx}       %
\usepackage{todonotes}
\usepackage{enumitem}
\usepackage{comment}
\usepackage{amsmath}
\usepackage{tabularx}
\usepackage{subcaption}

\title{SMART: Tuning a symbolic music generation system with an audio domain aesthetic reward}

\author{%
  Nicolas Jonason$^*$\\
  KTH Royal Institute of Technology\\
  \texttt{njona@kth.se} \\
  \And
  Luca Casini$^*$\\
  KTH Royal Institute of Technology\\
  \texttt{casini@kth.se} \\
  \And
  Bob L. T. Sturm\\
  KTH Royal Institute of Technology\\
  \texttt{bobs@kth.se} \\
}

\begin{document}

\maketitle
\footnotetext[1]{$^*$Equal contribution.}

\begin{abstract}
Recent work has proposed training machine learning models to predict aesthetic ratings for music audio.
Our work explores whether such models can be used to finetune a symbolic music generation system with reinforcement learning, and what effect this has on the system outputs.
To test this, we use group relative policy optimization to finetune a piano MIDI model with Meta Audiobox Aesthetics ratings of audio-rendered outputs as the reward.
We find that this optimization has effects on multiple low-level features of the generated outputs, and improves the average subjective ratings in a preliminary listening study with $14$ participants.
We also find that over-optimization dramatically reduces diversity of model outputs.
Several listening examples can be found here:
\href{https://anonymous-submission-199999999.github.io/SMART-demo/}{https://anonymous-submission-199999999.github.io/SMART-demo/}.

\end{abstract}

\section{Introduction}
\label{introduction}

Systems that predict human aesthetic preferences of music are gaining traction \citep{tjandra2025meta, cideron2024musicrl}. 
Unlike other evaluation methods that compare generated outputs against a reference set \citep{kilgour2018fr}, aesthetic preference models directly predict human preference ratings from a single audio input. 
This is achieved by collecting human ratings for various music stimuli and training a statistical model to predict average ratings for new examples. 
Despite the limitation that predicting the average rating disregards the subjectivity of musical preferences---treating aesthetic judgment as universal rather than listener-dependent---such models have seen adoption as heuristics for dataset filtering \citep{tjandra2025meta}, system evaluation \citep{mureka} and as reward for finetuning music generation models with reinforcement learning (RL) \citep{cideron2024musicrl}. While such aesthetic preference models have been developed for music audio, no such models have yet been developed for symbolic representations of music. 

Symbolic representations, such as MIDI, abstract musical elements into symbols that have to be rendered into audio in order to be listened to. 
Applying an audio aesthetic preference model to the development of a symbolic music generation model can be achieved in a few ways.
The most obvious way is to pre-render a symbolic music dataset with a symbolic-to-audio rendering pipeline to obtain aesthetic ratings for the symbolic music. These ratings can then be used as a heuristic for data filtering or as labels for distilling the audio preference model into a symbolic music preference model.
However, if the symbolic dataset is large this requires significant upfront computation and making decisions upfront about the rendering pipeline which ultimately limits the flexibility of this approach.

We explore a different option; using symbolic-to-audio rendering and an audio preference model to finetune a pretrained symbolic music model with reinforcement learning.  
We call this intervention \textbf{SMART}: \textbf{S}ymbolic \textbf{M}usic \textbf{A}udio \textbf{R}eward \textbf{T}uning. 
This brings us to our research question:
\emph{What are the effects of using an audio-domain aesthetic preference model as a reward for finetuning a symbolic music system?}.

We operate in the domain of piano music as there is a well established tradition of generating piano music \citep{performance-rnn-2017, huang2018improved, hadjeres2021piano, hawthorne2018transformer}. Additionally, piano music lends itself to be represented with MIDI and can easily be rendered into audio with somewhat high quality \citep{colton}.

The rest of the paper is structured as follows:
First we present related work on RL applications to music generation in Section \ref{sec:related}. 
Section \ref{sec:approach} presents our method and its constituent parts.
In Sec. \ref{sec:experiments} we present our experiments and results.
Section \ref{sec:discussion} discusses our results and present some additional findings.
Finally, Sec. \ref{sec:conclusions} concludes the paper and suggests avenues for future work.

\section{Related Work}
\label{sec:related}

Past work has used RL for optimizing music generation systems with a variety of reward signals.
\textit{RL-Tuner} \citep{jaques2017tuning}, \textit{Bach2Bach} \citep{kotecha2018bach2bach} and \textit{jaki} \citep{bruford2020jaki}, use Deep Q-Learning (DQN) methods for fine-tuning Long Short Time Memory models with symbolic reward functions. 
These methods operate entirely in symbolic space, without audio rendering in the reward loop. 
\textit{RaveForce} \citep{lan2019raveforce} also uses DQN, and incorporates audio rendering into the RL pipeline by synthesizing MIDI outputs and computing perceptual or spectral rewards on the resulting waveforms, enabling optimization of timbral or rhythmic features beyond symbolic metrics.
More recently, \textit{MusicRL} \citep{cideron2024musicrl} learns a reward model trained on more than $300,000$ human comparisons and fine-tunes a text-to-music model using Proximal Policy Optimization \citep{schulman2017proximal}  aligning generation quality with perceptual preferences. 
In the symbolic domain, \textit{NotaGen} \citep{wang2025notagen} features an RL fine-tuning phase based on Direct Policy Optimization \citep{rafailov2023direct}, aimed at refining music generation with respect to structured prompts, including composer, period, and instrumentation. 
The reward signal is obtained by measuring distances from a symbolic reference dataset in the \textit{CLAMP-3} embedding space \citep{wu2025clamp}.
What contrasts our work with earlier work is that we apply a audio aesthetic preference model as a reward for tuning a symbolic music system.

\section{Method}
\label{sec:approach}
\begin{figure}[ht] %
    \centering\includegraphics[width=1.0\linewidth]{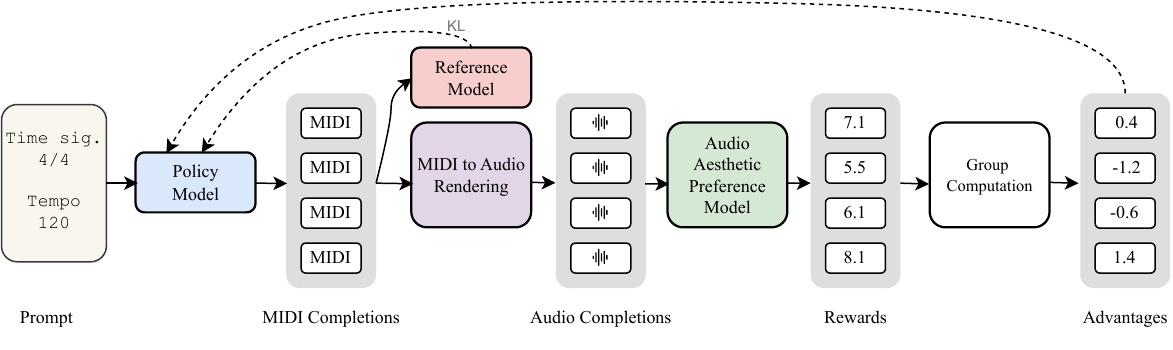}
    \caption{\textbf{Overview of the SMART training setup.} First, the policy and reference model are initialized from the pretrained base model. The reference model's weights are frozen. Then, each iteration, a prompt is passed to the policy model which generates MIDIs. The MIDIs are then rendered into audios which are assigned rewards by the aesthetic preference model. These rewards are then used to compute the group relative advantages which are then used to update the policy model. This optimization is regularized by an additional KL loss term which prevents the policy model from straying too far away from the reference model. This figure is adapted from \cite{shao2024deepseekmath}.}
    \label{fig:overview}
\end{figure}

In this section we explain how we trained the base model and the components involved in SMART training. 
Figure \ref{fig:overview} provides an explanation of the SMART training setup.

\subsection{Base model training}\label{sec:models}

We start by training a base model on a large dataset of MIDI files containing piano music.
We use a causal Transformer model using the Microsoft Phi 3 architecture \citep{abdin2024phi} as implemented in the Huggingface Transformers library.\footnote{\url{https://huggingface.co/docs/transformers/en/model_doc/phi3}, last accessed Apr. 21 2025.}
Our model has $4$ layers and a hidden size of $512$, with $8$ attention heads.
The model is trained on piano data from the \textit{MetaMIDI} dataset \citep{ens2021building}. 
Specifically, we filtered out any MIDI file that does not contain a program number belonging to the piano group in the \textit{General MIDI} specification (programs 1-8). 
Of these, only the piano tracks are kept and merged into a single track, removing duplicated notes. 
The MIDI files are tokenized using the \textit{REMI+} tokenizer \citep{rutte2023figaro} as implemented in the \textit{MidiTok} library \citep{fradet2023miditok}
After tokenization, additional filtering removes tracks with more than $300$ notes per bar and more than $20\%$ of empty bars, resulting in a total of $216,850$ files.
$10\%$ of the dataset is left as a holdout test set, while the remainder is used for test and validation with a 90-10 ratio.
During training we take random crops with a length of up to $512$ tokens, padding with special "PAD" tokens when necessary.

\subsection{Group Relative Preference Optimization}
\label{sec:grpo}

We employ \textit{Group Relative Preference Optimization (GRPO)} \citep{shao2024deepseekmath} as our RL algorithm, using the implementation from the TRL library.\footnote{\url{https://huggingface.co/docs/trl/main/en/grpo_trainer}, last accessed Apr. 21 2025}
Equation \ref{eq:grpo} shows the GRPO objective.
\begin{equation}\label{eq:grpo}
    \mathcal{L}_{GRPO}(\theta) = -\frac{1}{G} \sum_{i=1}^{G} \sum_{t=1}^{|o_i|} \left[ \frac{\pi_{\theta}(o_{i,t} \mid q, o_{i,<t})}{\pi_{\theta_{old}}(o_{i,t} \mid q, o_{i,<t})} \hat{A}_{i,t} - \beta D_{KL} \left[ \pi_{\theta} || \pi_{ref} \right] \right]
\end{equation}
$G$ is the number of groups, with each group containing multiple \emph{completions} $o_i$ of a specific \emph{prompt}.
$\pi$ and $\pi_{old}$ refer to the current and previous policy (model).
The advantage $\hat{A}_{i,t} = \frac{r_i + mean(r)}{std(r)}$ represent the group-normalized reward for each completion. 
This reward maximization term is regularized by the Kullback–Leibler (KL) divergence between a prediction, scaled by a parameter $\beta$. 
This second term is used to mitigate over-optimization toward the reward by penalizing completions which are unlikely according to the reference model $\pi_{ref}$.

\subsection{Audio Aesthetic Preference Model}
\label{sec:MAA}

\textit{Meta Audiobox Aesthetics} (MAA) is a model created for the assessing the quality of audio, speech and music \citep{tjandra2025meta}.
It is trained on $10$s fragments of audio, which human raters scored on a 10-point scale across four different aesthetic attributes described by \citet{tjandra2025meta} as follows:
\begin{description}[leftmargin=0.0cm] 
    \item[Production Quality] ``Focuses on the technical aspects of quality instead of subjective quality. Aspects including clarity \& fidelity, dynamics, frequencies and spatialization of the audio'';
    \item[Production Complexity] ``Focuses on the complexity of an audio scene, measured by number of audio components'';
    \item[Content Enjoyment] ``Focuses on the subject quality of an audio piece. It’s a more open-ended axis, some aspects might includes emotional impact, artistic skill, artistic expression, as well as subjective    experience, etc'';
    \item[Content Usefulness] ``Also a subjective axis, evaluating the likelihood of leveraging the audio as source material for content creation.''
\end{description}
MAA has some limitations. First, the 10 second receptive field means it can not take into account 
patterns that are longer than 10 seconds, such as the repetition and development 
of musical ideas in a coherent structure.
Secondly, music collection used, and the cultural background and musical training of the raters, 
are not disclosed by \citet{tjandra2025meta}.

Table \ref{tab:audiobox_real} gives some examples of MAA ratings of real piano performances as well as white noise and silence to provide more context on what these ratings may mean.
(Table \ref{tab:performance_urls} in the Appendix provides the URLs for the recordings.)
For our experiments we choose \emph{content enjoyment} as the reward because we believe that is the most appropriate for a music generation system.

\begin{table}[htb]
\caption{Meta Audiobox Aesthetics ratings for recordings of 10-second excerpts of various piano performances. We also include the ratings given to silence and white noise. CE: Content Enjoyment. CU: Content Usefulness. 
PC: Production Complexity. PQ: Production Quality}
\label{tab:audiobox_real}
\begin{tabularx}{\textwidth}{Xllll|l}
\toprule
Audio & CE & CU & PC & PQ & Mean \\ 
\midrule
Oscar Peterson - Someone to Watch Over Me (Live) & 7.48 & 7.76 & 3.94 & 7.70 & 6.72 \\
Ludovico Einaudi - Nuvole Bianche                & 7.41 & 7.91 & 3.47 & 8.01 & 6.70 \\
Chopin - Waltz Op.69 No.2 (Ashkenazy)            & 7.60 & 7.92 & 3.50 & 7.50 & 6.63 \\
Conlon Nancarrow - Study for Player Piano No. 21 & 6.40 & 7.71 & 3.95 & 7.88 & 6.49 \\
Philip Glass - Etude No.6 (Yuja Wang)            & 6.88 & 7.61 & 3.99 & 7.43 & 6.48 \\
Keith Jarrett - Solar                            & 7.12 & 7.35 & 3.90 & 7.48 & 6.46 \\
Brad Mehldau - My Favorite Things (Live)         & 7.24 & 7.63 & 3.36 & 7.28 & 6.38 \\
Lili Boulanger - Prelude in B Major              & 6.80 & 7.64 & 3.38 & 7.49 & 6.33 \\
Hiromi Uehara - Blackbird (Live)                 & 6.27 & 7.06 & 3.63 & 7.24 & 6.05 \\
John Cage - Sonata for Prepared Piano No.1       & 5.96 & 7.8 & 2.62 & 7.75 & 6.03 \\
Mozart - Piano Sonata No.16 K.545 (Ingrid Haebler) & 6.95 & 7.33 & 3.03 & 6.71 & 6.01 \\
\midrule
Silence                                          & 3.24 & 6.45 & 1.7 & 6.74 & 4.53 \\
White Noise                                      & 2.37 & 4.65 & 1.99 & 4.79 & 3.45 \\ 
\bottomrule
\end{tabularx}
\end{table}

\subsection{MIDI to Audio Rendering}\label{sec:rendering}

We render MIDI into audio using the \textit{TinySoundfont}\footnote{\url{https://github.com/schellingb/TinySoundFont}, last accessed Apr. 21 2025} soundfont renderer. Soundfonts is a format for sample-based synthesis commonly used for rendering MIDI files. Table \ref{tab:soundfonts} shows the average for the four different ratings produced by MAA computed over a batch of $10$ samples from the pretraining dataset rendered with five different publicly available soundfonts. 
(The URLs of those soundfonts are provided in Table \ref{tab:soundfonturls} in the Appendix.)
For the experiments described in this paper we use the \texttt{Yamaha C5 Salamander-JNv5.1} soundfont.

\begin{table}[htb]
    \caption{ Audiobox Aesthetics rating for five publicly available soundfonts.}
    \label{tab:soundfonts}
    \begin{tabularx}{\linewidth}{Xrrrrr}
    \toprule
    Soundfont & \texttt{musescore} & \texttt{fluidr3} & \texttt{grandeur} & \texttt{sgm} & \texttt{yamaha} \\
    \midrule
    Content Enjoyment & 5.423 & 5.347 & 6.385 & 6.602 & 6.530 \\
    Content Usefulness & 7.487 & 7.408 & 7.693 & 7.716 & 7.705 \\
    Production Complexity & 2.352 & 2.355 & 2.376 & 2.477 & 2.453 \\
    Production Quality & 7.323 & 7.817 & 7.884 & 7.695 & 7.883 \\ \midrule
    Average & 5.646 & 5.732 & 6.085 & 6.123 & 6.143 \\
    \bottomrule
    \end{tabularx}
\end{table}

\subsection{SMART training details}
\label{sec:training-details}
\begin{figure}[h]
    \centering
    \includegraphics[width=0.5\textwidth]{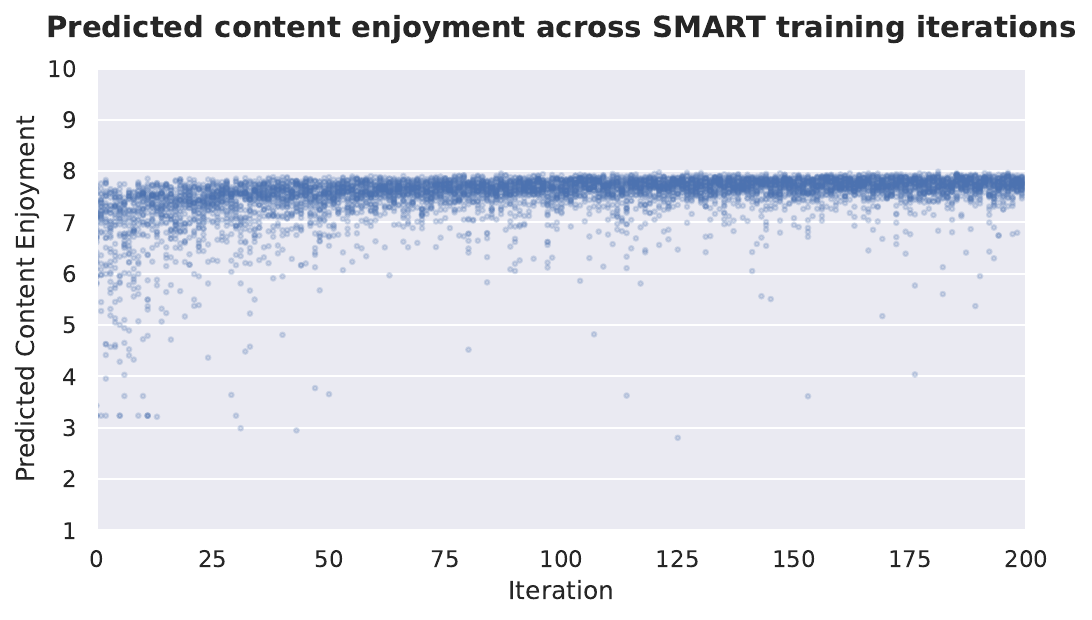}
    \caption{Predicted content enjoyment rating from MAA across SMART training iterations.}
    \label{fig:training}
\end{figure}

Each iteration,  we procedurally generate $8$ prompts formatted according to the REMI+ syntax \citep{rutte2023figaro}, including a token indicating the start of a new bar, a random time signature token and a random tempo token. 
For each prompt, we generate $8$ completions, totaling $64$ completions per batch. 
All completions were sampled with temperature $1.0$. 
We cropped all audio to $10$ seconds, the receptive field of MAA.
We set the KL divergence regularization parameter $\beta=0.04$ and employed a linear learning rate schedule starting at $1 \times 10^{-4}$ and decaying to $0$ over $200$ iterations.
Training was conducted on a single \textit{NVIDIA GeForce RTX 3090} GPU and completed in $50$ minutes. Figure \ref{fig:training}
shows the reward improving over the training iterations.%

\section{Results}
\label{sec:experiments}
In this section we will illustrate the results from multiple analyses conducted on the outputs of our models. We compare the base model outputs and SMART model outputs by looking at MAA ratings, low level features, and subjective ratings from a small listening test. We also invite the reader to listen to the audio examples presented at \href{https://anonymous-submission-199999999.github.io/SMART-demo/}{https://anonymous-submission-199999999.github.io/SMART-demo/}.

\subsection{Comparison of Meta Audiobox Aesthetics ratings}
\begin{figure}[htb]
    \centering
    \includegraphics[width=1.0\linewidth]{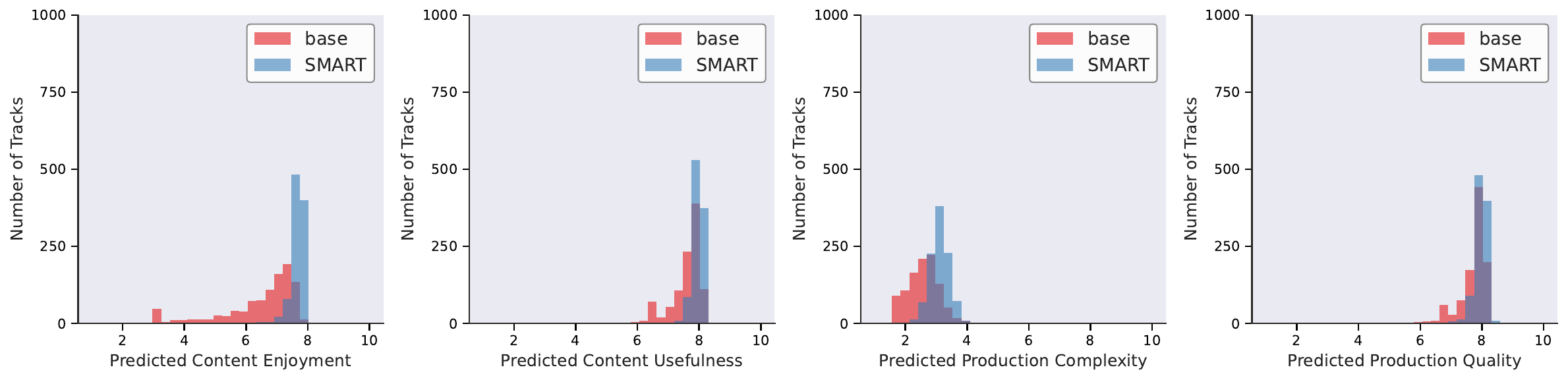}
    \caption{\textbf{Optimizing for predicted content enjoyment also increases the other ratings.} MAA ratings of 1000 generations with random procedural prompts for the piano model pre and post audio reward optimization.}
    \label{fig:audiobox}
\end{figure}

Figure \ref{fig:audiobox} shows that SMART training is able to optimize the Content Enjoyment reward. 
We also notice increases in the MAA ratings we did not optimize for. %

\subsection{MIDI-level features}

\begin{figure}[htb]
    \centering
    \includegraphics[width=1.0\linewidth]{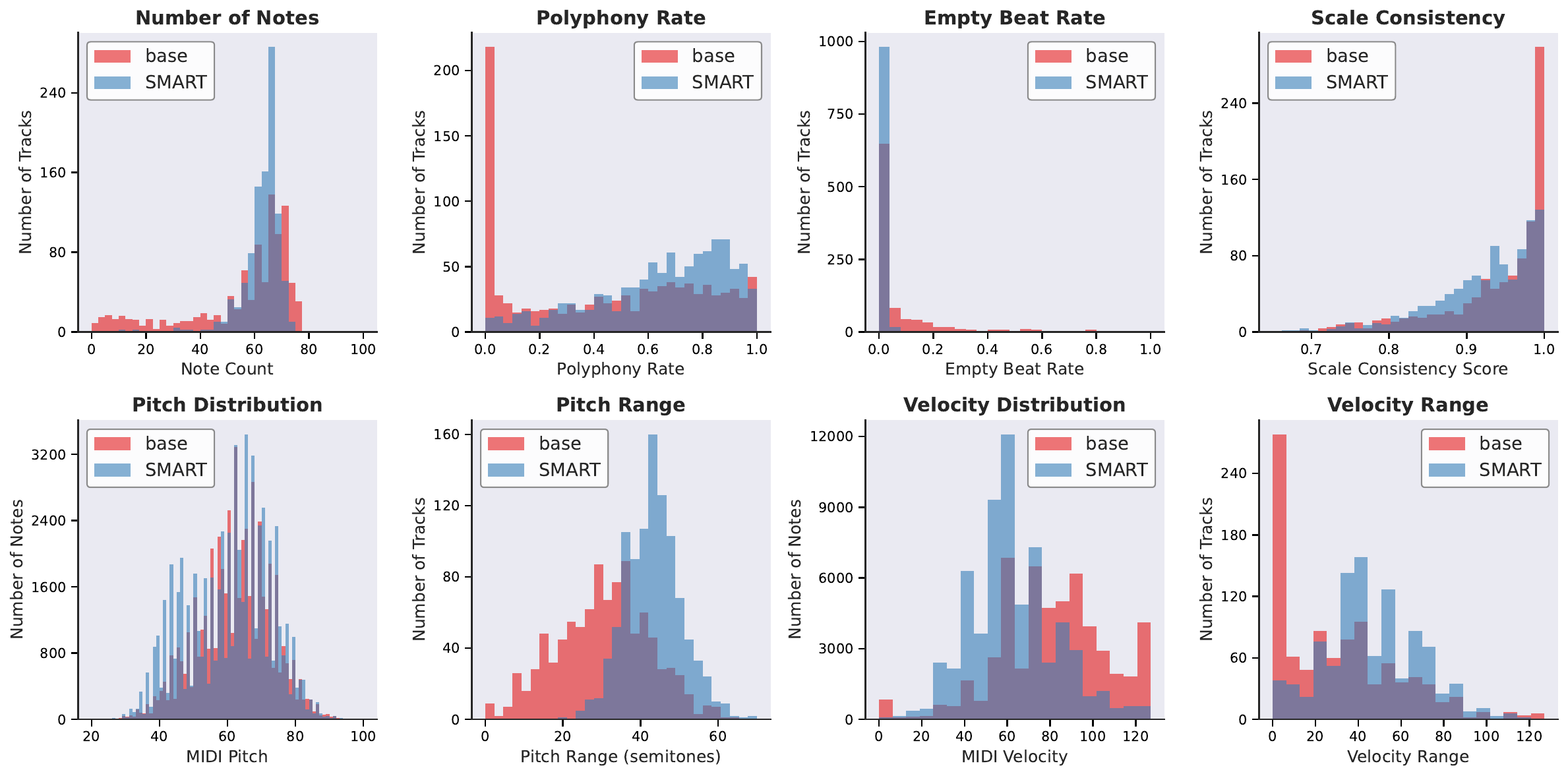}
    \caption{
    \textbf{Optimizing towards the aesthetic reward affects distribution of low-level features in model outputs.}
    Histograms of various track and note features from 1000 generations with random procedural prompts for the piano model before and after SMART training.}
    \label{fig:lowlevel}
\end{figure}

We investigate the effect of our intervention by comparing MIDI-level features of the outputs from both the base model and SMART model.
Using $1000$ procedurally, we generate one completion from the base model and one from the SMART model, producing $2000$ MIDI files.
Using {MusPy} library \citep{dong2020muspy} we compute $8$ features whose histograms are shown in Figure \ref{fig:lowlevel}.

\begin{description}[leftmargin=0.0cm]
    \item[Number of notes] refers to the count of note events in the midi file. We see that the base model's outputs sometimes contain few notes. This is not the case for the post-intervention model whose note counts are concentrated around $70$.
    \item[Polyphony rate] is the proportion of time steps where multiple pitches are active. We see that the post-intervention model uses more polyphony.
    \item[Empty-beat rate] is the ratio of beats that contain no note event. While the base model's outputs contains some empty beats, these are more rare in the post-intervention model.
    \item[Pitch distribution:] This shows the count of each individual MIDI pitch. Compared to the base model, the SMART model has more pitches in the lower octaves.
    \item[Pitch range] refers to the distance in semitones between the lowest and highest pitch in a track. The SMART model tends to use a larger pitch range.
    \item[Scale consistency:] This is the highest ratio of notes that belong to a certain scale, tested against all major and minor scales.
    The distribution appears to have a heavier tail after SMART training, suggesting increased use of chromaticism.
    \item[Velocity distribution:] This refers to the count of individual velocity levels (the tokenizer uses $20$ bins).
    Compared to the base model, the SMART model outputs tend to have lower velocities. 
    \item[Velocity range:] This is the distance between the lowest and highest velocity value in a track. The base model shows a zero-inflated distribution, suggesting no dynamic range is present in some of its outputs. This is more rare in the SMART model whose outputs tend include a larger range of velocity values.
\end{description}

\subsection{Listening Study}
\begin{figure}[h]
    \centering
    \begin{subfigure}[t]{0.4\textwidth}
        \centering
        \includegraphics[height=4cm]{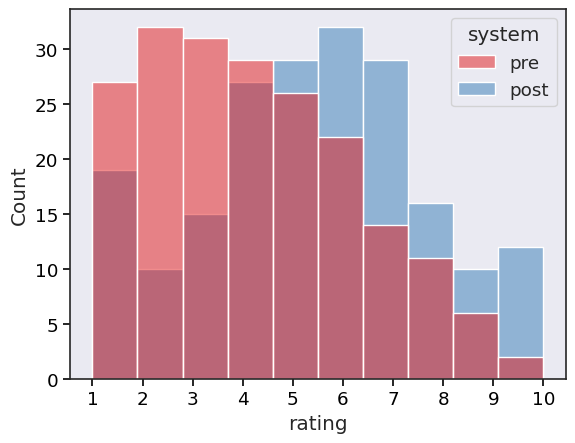}
    \end{subfigure}%
    ~ 
    \begin{subfigure}[t]{0.6\textwidth}
        \centering
        \includegraphics[height=4cm]{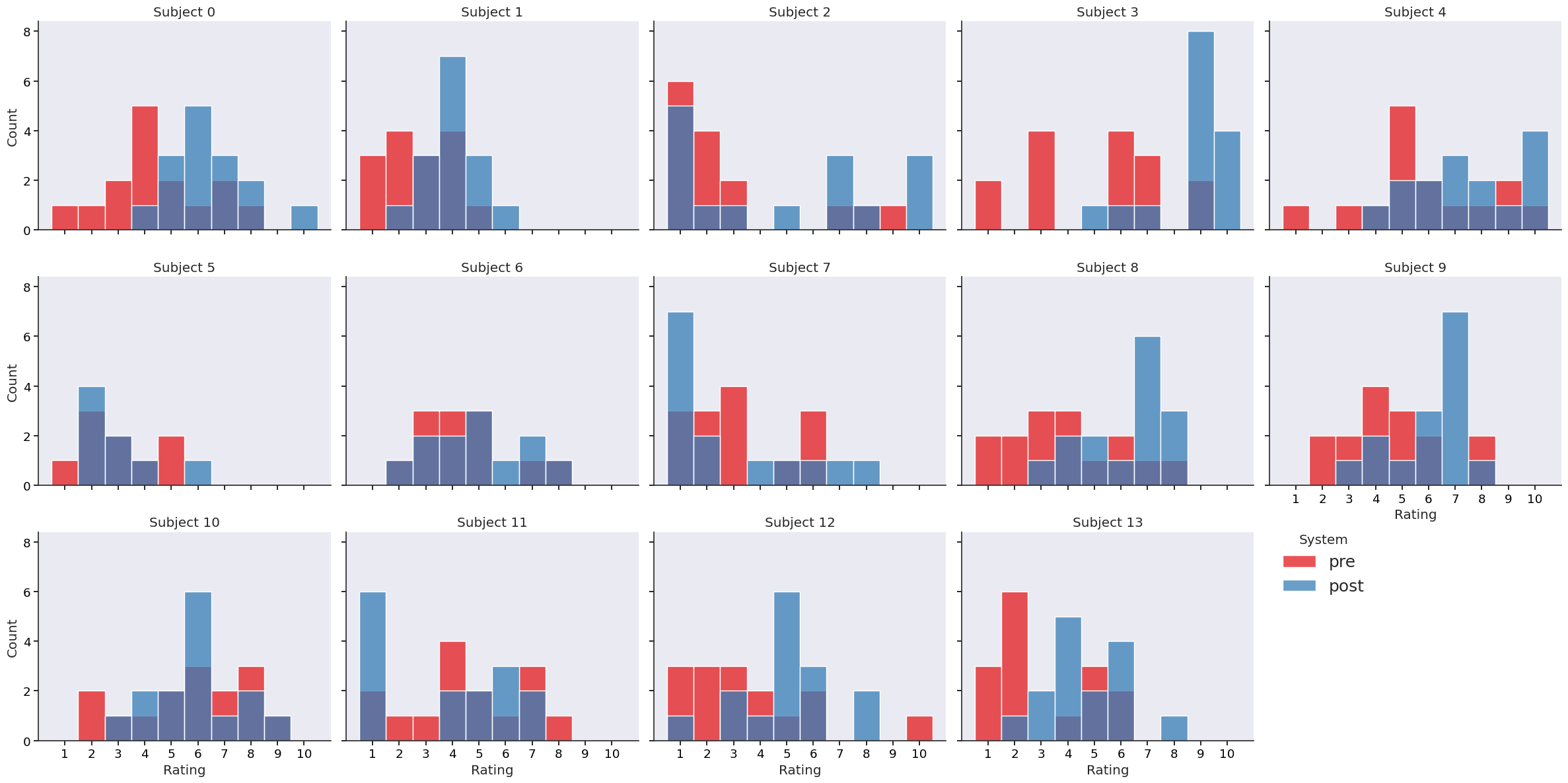}
    \end{subfigure}
        \caption{
    \textbf{On the left:} Overall distribution of the ratings. Red indicates the base model, blue indicates the post-intervention model.
    \textbf{On the right:} Distribution of the ratings from each subject.
    }
    \label{fig:plot}
\end{figure}

We conduct a small listening study, collecting ratings from $14$ subjects recruited through convenience sampling from researchers in a computer science department in a European university.
Before the test starts, each participant is prompted with the following message: 
\textit{``
You will listen to several audio samples and rate their enjoyability``}.
In each trial, a participant listens to a $\sim10$s music excerpt, and is then asked, 
\textit{``How much do you enjoy this audio? Rate from 1 to 10.''}
Each subject rated $30$ excerpts in total, half generated by the base model and the other half from the SMART model, from the same set of 15 random prompts. The excerpts were presented in random order without labels.
For the last three participants, a page was added to the welcome screen stating the project funding the listening test.
At the end of the study we obtain $407$ ratings, as one participant failed to finish the test.
From this, we remove $8$ observations corresponding to $4$ prompts that produced silent outputs from the base model.
The total number of observation is thus $399$.
Figure \ref{fig:plot} shows the distribution of ratings overall as well as per participant.

To further consolidate the results, we fit a linear mixed effect model 
\begin{equation}\label{eq:lmer}
    \mathbf{y} = \boldsymbol{X\beta} + \boldsymbol{Zu} + \boldsymbol{\varepsilon}
\end{equation}
where $\mathbf{y}$ are the ratings, $\boldsymbol{X\beta}$ represents the fixed effects predictors and coefficients, $\boldsymbol{Zu}$ represent the random effects and $\boldsymbol{\varepsilon}$ are the residuals.
The choice of model is dictated by the necessity to control for subject depended effects in the model.
In our case the only fixed effect is the SMART intervention.
The random effect include a random intercept and slope for the stimulus group and for the subject group. 
We implemented the model in R using the \texttt{lmer} package. 
Table \ref{tab:lmer} in the Appendix summarizes the results.
We see a significant ($p = 0.002$) positive effect of the SMART intervention on participants' ratings. 
On average, ratings increased by approximately $1.22$ points in the post condition compared to the baseline.

\section{Discussion}\label{sec:discussion}

\subsection{How does SMART affect the model outputs?}

The results indicate that the SMART intervention has a number effects on model behaviour.
Post SMART model outputs have larger numbers of notes, have increased rate of polyphony, reduces the empty beat rate to zero, uses a wider pitch range, and use a wider set of velocities. Additionally, the velocities of the post intervention model are lower. 
Comparing the outputs in our own listening of the outputs between the models, the most apparent changes are that the number of pauses are less frequent and that it has more dynamics.
These factors could explain why the listeners find the SMART model's outputs more enjoyable on average.
One interesting observation is that the post-intervention models outputs obtains MAA ratings (Figure \ref{fig:audiobox}) that often exceed the MAA ratings from real performances in Table \ref{tab:audiobox_real}. This should not be interpreted as the model outputs being better than the real performances but rather shows the limitations of MAA as a proxy for human subjective experience.

\subsection{What happens if we optimize more aggressively?}

\begin{figure}[htb]
    \centering
    \includegraphics[width=\linewidth]{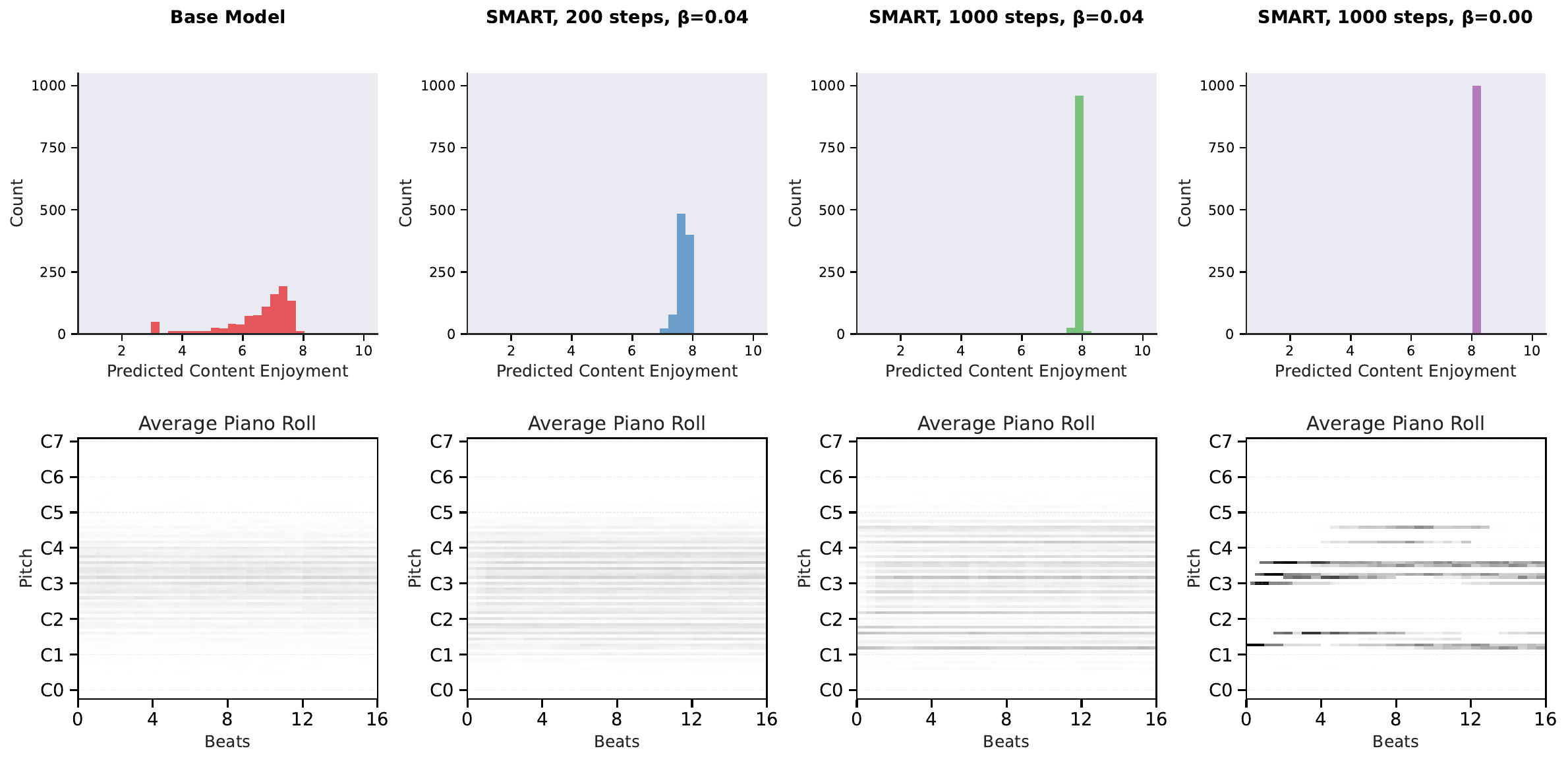}
    \caption{
    \textbf{Aggressive optimization of models towards the aesthetic reward results in high rewards but destroys diversity in outputs.}
    The distribution of Content Enjoyment scores (top) and average piano rolls from first 16 beats (bottom) computed on $1000$ samples from the base model and three SMART models with progressively more aggressive optimization settings.
    }
    \label{fig:overoptimization}
\end{figure}

A known phenomenon in reinforcement learning is that optimization towards a proxy reward past a critical point hurts performance on the true objective \citep{karwowski2024goodharts}. 
To test how this would manifest itself in our setting, we ran a SMART optimization over $1000$ iterations instead of $200$. To push the over-optimization further, we also 
set $\beta=0.0$ and ran the optimization for $1000$ steps.
Figure \ref{fig:overoptimization} shows the rewards and average piano roll over $1000$ outputs from random prompts. 
When optimizing more aggressively, we observe that the model's outputs obtain higher ratings from MAA, but also that the average piano rolls show more distinct patterns, indicating reduced diversity in the outputs. The loss in diversity is consistent with earlier findings in the field of reinforcement from human feedback for large language language models \citep{kirk2024understanding}.

\subsection{What if we use prompts from the pretraining set instead?}
    
\begin{figure}
    \includegraphics[width=\linewidth]{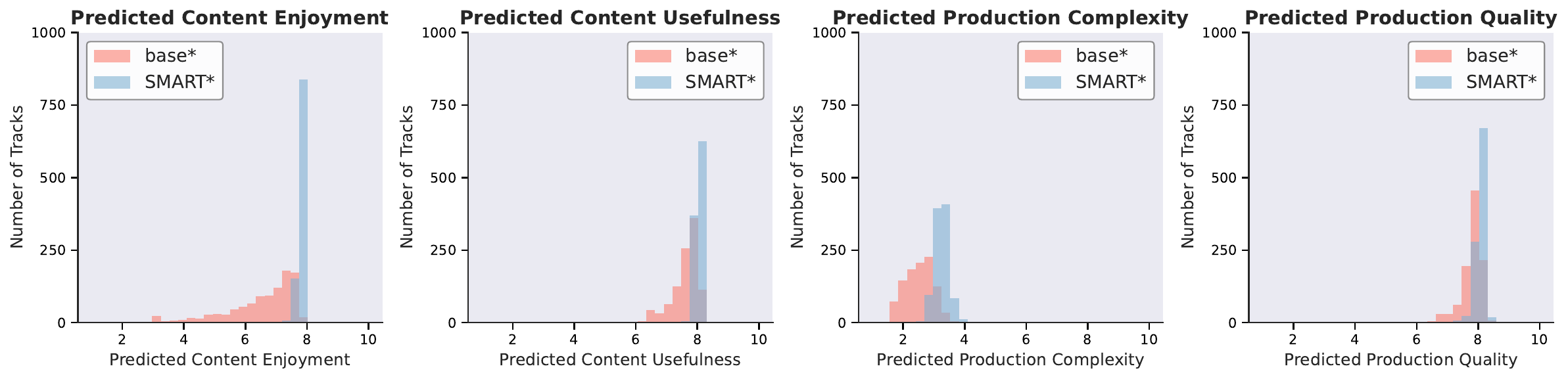}
    \includegraphics[width=\linewidth]{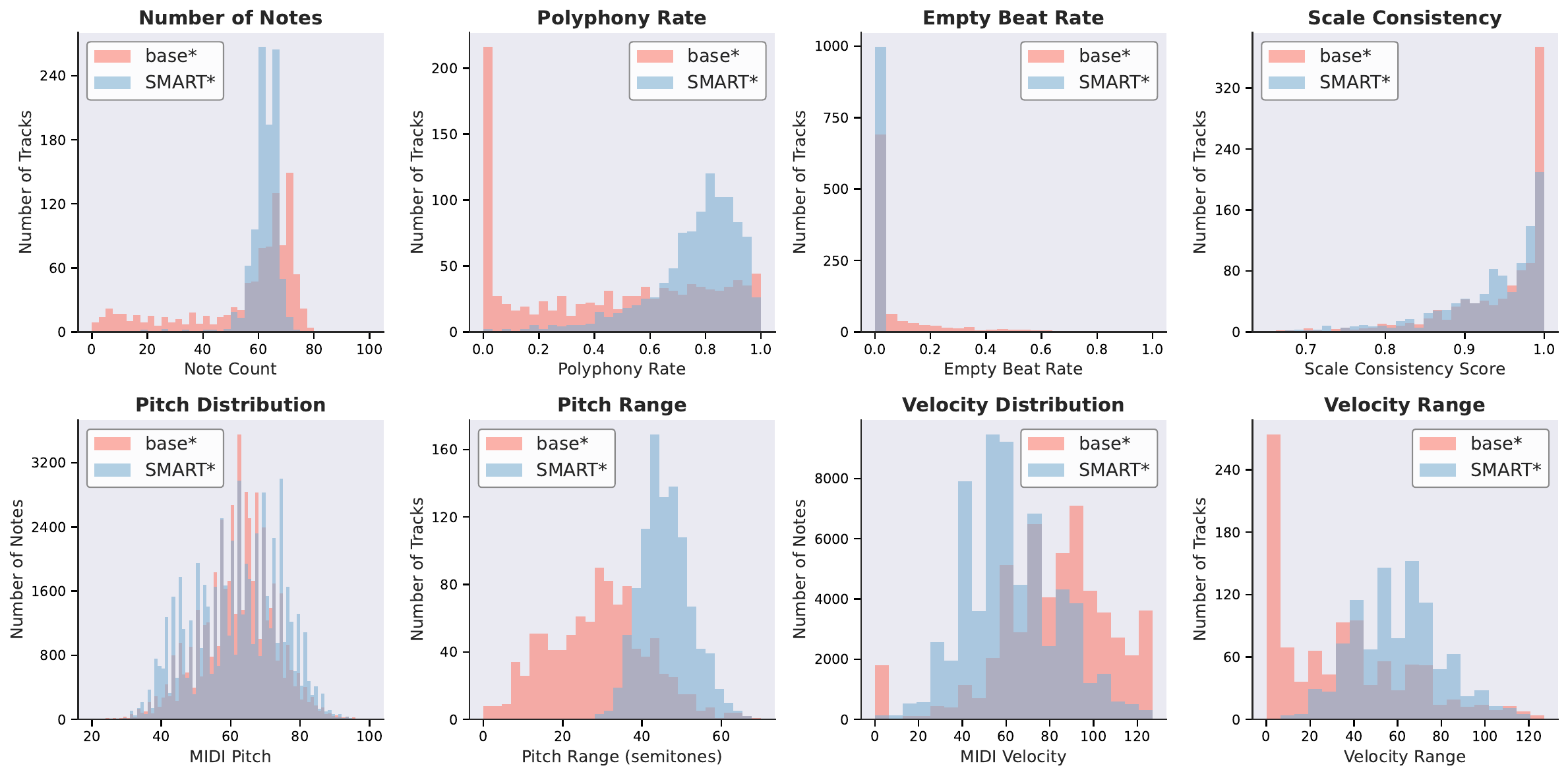}
    \centering
    \caption{MAA ratings and MIDI features for $1000$ outputs from the base model and after $200$ iterations of SMART training. *\emph{Prompts extracted from the pretraining dataset}.}
    \label{fig:prompts-from-dataset}
\end{figure}

One consequence of the procedural prompts used during SMART training is that we induce a domain shift with respect to the pretraining data used. Tempos and time signature which are rare in the pretraining set are more represented in the procedural prompts. We therefore wanted to understand if using prompts collected from the pretraining dataset would result in a different overall outcome. We therefore ran another SMART training run with the same settings as in section \ref{sec:training-details}
except for the prompts being sampled from the training split of the pretraining dataset. We then sampled $1000$ prompts from the test split of the pretraining data and computed MAA scores and MIDI features on the outputs of the base model and the resulting SMART model. As seen in Figure \ref{fig:prompts-from-dataset}, we observe similar increases in the MAA ratings and changes in MIDI features in this setting as in the procedural prompt setting.

\section{Conclusions}\label{sec:conclusions}
We investigate the effects of tuning a piano symbolic music generation model with a audio domain aesthetic reward. We find multiple notable differences between the outputs before and after this intervention, including more notes, less pauses, and more dynamics. We also find that for listeners in our small sample of 14 academics from a computer science department, the intervention results in an increase in enjoyability scores. However, testing the effect of the intervention on listeners from a wider demographic in a more formal study is necessary to see the extent to which the same trend holds in the general population. Future work will test the proposed approach with different audio aesthetic preference models and extend the work to the multi-instrument setting.

\section*{Ethics Statement}

Our pretraining dataset consists of a subset of the MetaMIDI dataset \citep{ens2021building} which was collected by scraping publically available websites. As stated in subsection \ref{sec:MAA}, MAA has some limitations which in turns has implications on the our models. First, the lack of information about the raters and music used for collecting ratings makes it hard to ascertain what cultural biases are embedded in the MAA model. Further, by focusing on predicting the average rating of music, MAA is unable to take into account the fact that music preferences are not universal. We hope that future work in this area will produce models which account for differences across human preferences. Considering the wider societal implications of our work, we can not rule out a future where AI technology reduces the demand for human artists. By working in the symbolic music domain and not targeting ready-for-consumption full music tracks, we hope that our work can aid the development of tools that serve artists rather than replacing them. Total carbon dioxide emissions associated with the compute in this project are estimated to be 9.07 kgCO$_2$eq as estimated by the ML CO2 Impact calculator \cite{lacoste2019quantifying}.

\section*{Acknowledgements}
This work is an outcome of a project that has received funding from the European Research Council under the European Union’s Horizon 2020 research and innovation program (MUSAiC, Grant agreement No. 864189).

\bibliographystyle{apalike}   
\bibliography{references}  

\section{Appendix}
\begin{table}[htbp]
\caption{Soundfonts used in experiments}
\label{tab:soundfonturls}
\small
\centering
\begin{tabularx}{\linewidth}{llX}
\toprule
\textbf{Short Name} & \textbf{Full Name} & \textbf{URL} \\
\midrule
musescore & MuseScore General & \url{https://musescore.org/en/handbook/3/soundfonts-and-sfz-files} \\
\hline
fluidr3 & FluidR3 GM & \url{https://musescore.org/en/handbook/3/soundfonts-and-sfz-files} \\
\hline
grandeur & The Grandeur D & \url{https://musical-artifacts.com/artifacts/2992} \\
\hline
sgm & SGM-V2.01-XG-2.04 & \url{https://archive.org/details/SGM-V2.01} \\
\hline
yamaha & Yamaha-C5-Salamander-JNv5\_1 & \url{https://sites.google.com/site/soundfonts4u/} \\
\bottomrule
\end{tabularx}
\end{table}

\begin{table}[htb]
\caption{URLs for Piano Performances Referenced in Table~\ref{tab:audiobox_real}}
\label{tab:performance_urls}
\centering
\small
\begin{tabularx}{\linewidth}{lX}
\toprule
\textbf{Performance} & \textbf{URL} \\
\midrule
Chopin - Waltz Op.69 No.2 (Ashkenazy) & \url{https://www.youtube.com/watch?v=cxG-kOTMgaA} \\
Ludovico Einaudi - Nuvole Bianche & \url{https://www.youtube.com/watch?v=CQ8zglIXZi8} \\
Mozart - Piano Sonata No.16 K.545 (Ingrid Haebler) & \url{https://www.youtube.com/watch?v=I_AX4R-d29o} \\
Philip Glass - Etude No.6 (Yuja Wang) & \url{https://www.youtube.com/watch?v=pjizB3A5g_0} \\
Oscar Peterson - Someone to Watch Over Me (Live) & \url{https://youtu.be/hZwfCKixNj4?si=mEcj-M6xN4LObYXl} \\
Brad Mehldau - My Favorite Things (Live) & \url{https://www.youtube.com/watch?v=rmSoK4cfMfI} \\
Keith Jarrett - Solar & \url{https://www.youtube.com/watch?v=zCKobOe7Cww} \\
Hiromi Uehara - Blackbird (Live) & \url{https://www.youtube.com/watch?v=DOsOLkQpGyA} \\
Lili Boulanger - Prelude in B Major & \url{https://www.youtube.com/watch?v=vyi3dTxffrM} \\
John Cage - Sonata for Prepared Piano No.1 & \url{https://www.youtube.com/watch?v=niu7jy5mB5k} \\
Conlon Nancarrow - Study for Player Piano No. 21 & \url{https://www.youtube.com/watch?v=hMnTWfZgunU} \\
\bottomrule
\end{tabularx}
\end{table}

\begin{table}[htb]
\footnotesize
\centering
\caption{Summary of Linear Mixed-Effects Model: \texttt{rating} $\sim$ \texttt{system} + (\texttt{system}$|$\texttt{id})}
\label{tab:lmer}
\begin{tabular}{@{}llrrrr@{}}

\toprule
\multicolumn{6}{c}{\textbf{Model Fit Statistics}} \\ 
\midrule
Number of observations &  &  &  &  & 399 \\
Groups (id) &  &  &  &  & 14 \\
Log-likelihood &  &  &  &  & -871.671 \\
AIC &  &  &  &  & 1755.343 \\
\midrule\midrule

\multicolumn{6}{c}{\textbf{Random Effects (by id)}} \\

\toprule
\textbf{Name} &  &  &  & \textbf{Variance} & \textbf{Std. Dev.} \\
\midrule
Intercept &  &  &  & 0.830 & 0.911 \\
SMART &  &  &  & 0.752 & 0.867 \\
Residual &  &  &  & 4.194 & 2.048 \\
\midrule
Correlation (Intercept, SMART) & \multicolumn{1}{r}{0.488} &  &  &  &  \\

\midrule\midrule
\multicolumn{6}{c}{\textbf{Fixed Effects}} \\

\toprule
\textbf{Term} & \textbf{Estimate} & \textbf{2.5\% CI} & \textbf{97.5\% CI} & \textbf{SE} & \textbf{p-value} \\
\midrule
Intercept & 4.075 & 3.520 & 4.631 & 0.284 & $<0.001^{***}$ \\
SMART & 1.221 & 0.613 & 1.828 & 0.310 & $0.002^{**}$ \\
\bottomrule
\end{tabular}
\end{table}

\end{document}